# Electron Condensation in a Jellium Coupled to a Finite Photonic Cavity

Pritha Mondal[1], Subham Kumar Saha[1], Awadhesh Narayan[1] and Anshu Pandey[1*]

[1]Solid State and Structural Chemistry Unit, Indian Institute of Science Bangalore 560012, India.

* Correspondence to: anshup@iisc.ac.in

*Abstract*

We describe the process of electron condensation into a localized state in a structureless jellium that is coupled to a finite cavity. It is shown that there exists a temperature $T_0$ below which electrons within the jellium localize. This process is driven by enhancement of correlations between the electrons that are coupled to the cavity.

*Main Text*

It is well known that cavities can serve to stabilize ground states of emitters that are coupled through enhancement in correlations of the ground states [1]. We consider the phenomena that arise when an infinite jellium is coupled to a finite cavity. This system is meant to serve as a rather simplified model for the Au-Ag family of materials that have been posited to show superconductivity at ambient temperature and pressure as well as suspected to exhibit a reduced free electron density [2,3]. A more detailed investigation of the properties of these materials in the context of quantum electrodynamically mediated electron condensation will be described elsewhere.

A finite sized cavity can couple to a dipole essentially by perturbing the local photonic density of state. In the strong coupling regime, this is approximately described by $\omega_c = \frac{\omega_0+\omega_m}{2} \pm \sqrt{\frac{F\gamma\Gamma}{4}}$ [4]. Here, $\omega_c$ are the real components of frequencies of the coupled system, $\omega_0$ and $\omega_m$ are the real components of frequencies associated with the dipole and the cavity respectively. $F$ is the Purcell factor of the cavity, $\gamma$ is its linewidth and $\Gamma$ is related to the spontaneous emission rate of the dipole. As evident, the effect leads to a splitting of the two interacting transitions, with an overall decrease of the resonance energy of the lower energy transition under favourable circumstances (i.e. when the detuning is sufficiently small).

A jellium with infinite spatial extent does not exhibit low energy dipole active transitions at its Fermi level. However, the restructuring of electrons into standing waves of similar spatial extent as the cavity allows them to potentially couple to the cavity modes. This carries an energy penalty $O(V^{-2/3})$ that becomes negligible for a sufficiently large volume $V$ of the cavity.

As described above, coupling to a cavity however changes the spectrum of electronic excitations that leads to the stabilization of the ground state of the many electron system. We illustrate this using a Moller-Plesset MP2 calculation [5]. Within an MP2 scheme, electron correlation energy is given by terms of the form $\frac{|I|^2}{E}$. Here $I$ represents the doubly excited electron integrals of the system. $E$ describes the two-electron excitation energy of the system. Due to the modified electromagnetic density of state, if the excitation energies of individual electrons are lowered on average by a factor of $\Delta$, the modified correlation energy contribution by each MP2 term is given by $\frac{|I|^2}{E-2\Delta}$. In the coupled system, each term thus leads to an excess correlation of $\frac{|I|^2}{E}\left(\frac{2\Delta}{E-2\Delta}\right)$. As evident, the lowest energy two-electron excitation make the most significant contributions to the excess correlation energy of this many electron system.

Consider the situation a large cavity of volume $V$ that is effectively coupled to $N$ jellium electrons; $n_b$ of these are bound electrons within the jellium while $n_f$ are free that may localize at sufficiently low temperatures, while $n_f^0$ is the minimum number of free electrons that are always present in the system due to reasons that are peculiar to certain systems. These examples will be described elsewhere. Thus, $N = n_b + n_f + n_f^0$. For this illustration, the cavity is taken to be large enough so that the confinement energy of the electrons is negligible. For $n_b$ electrons confined to a region of volume $V$, the associated confinement energy is $\sim P n_b^{5/3}\left[V^{-2/3} - V_0^{-2/3}\right]$ for $V < V_0$ and 0 for $V \geq V_0$. This estimate of relies on treating the effects of confinement on the electron levels as is done for a particle in a box. Similar approximations are applied for example in the case of semiconductor nanocrystals. Here $P$ is a mass dependent constant and $V_0$ is the localization volume in bulk. This becomes negligible for a cavity O(10 nm) in most cases.

To further simplify our discussion, we define $N' = N - n_f^0 = n_b + n_f$. If we limit ourselves to consider only the contributions to correlation energy that arise from the lowest energy excited determinants, we find that there are $O((n_b)^{4/3})$ such terms. This suggests an excess correlation energy $E \sim E_0 \left(\frac{2\Delta}{E - 2\Delta}\right)(n_b)^{4/3} = C(n_b)^{4/3}$, where $E_0$ is the MP2 correlation energy per electron for a free electron gas. Further, for a cavity that is not detuned from the lowest excitation energy of the many electron system, $\Delta \sim \sqrt{\frac{F\gamma\Gamma}{4}}$.

The electron localization process is shown schematically in Figure 1a. A fraction of the system's electrons can localize to couple to the cavity mode. Assuming a classical probability of the many electron system to sample one of the two configurations, $n_f = N' \exp\left(-\frac{C(n_b)^{4/3}}{kT}\right)$.

We therefore obtain the probability for an electron to be bound:

$$f_b = \frac{n_b}{N} = \frac{N'}{N}\left[1 - \exp\left(-\frac{C(f_b)^{\frac{4}{3}}}{kT}\right)\right] \sim \frac{N'}{N}\left[1 - \exp\left(-1.73\frac{T_0}{T}f_b^{4/3}\right)\right].$$

This form is suggestive of a phase transition in the system, where electrons coupled to the cavity localize below a certain critical temperature $T_0$. This equation has only a trivial solution for $T > T_0$. At $T = T_0$ this equation has a trivial solution as well as a finite one. For a certain temperature range with $T < T_0$ the equation has three possible solutions and again reverts to two solutions for 0 K. In each case we only admit the solution that corresponds to the lowest electronic energy. This physically tangible solution is presented in Figure 1b that shows the plot of $f_b$ as a function of temperature for $n_f^0 = 0$ (green, thick line) and $n_f^0 = \frac{N}{2}$ (blue thin line).

In conclusion, we describe the condensation of electrons due to enhanced correlations that occur when a system is coupled to a finite cavity. Our results are useful to understand the behaviour of certain materials such as the Au-Ag system where the disappearance of usual plasmon resonances has been observed. These aspects will be described in a subsequent note.

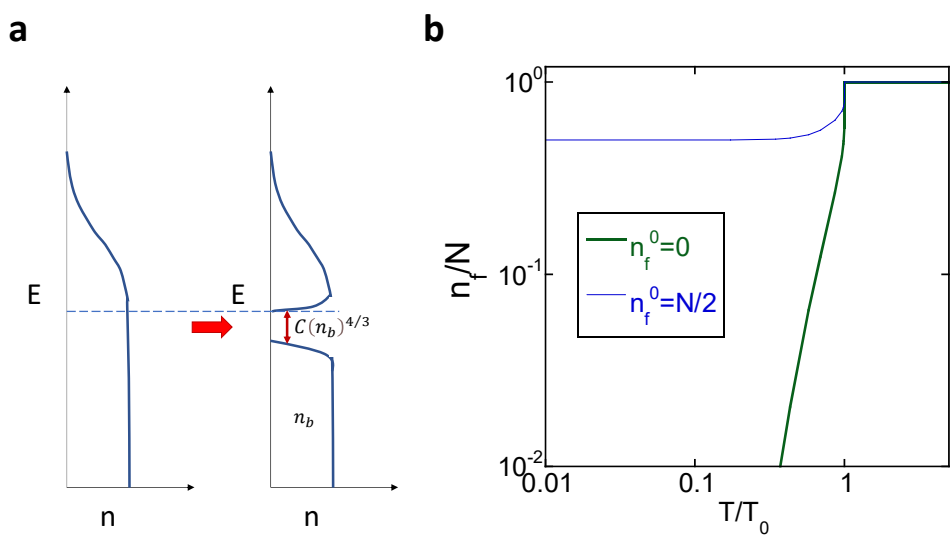

**Figure 1. a** Schematic of electron condensation in presence of a cavity. **b** Probability of an electron to be free in region of cavity as a function of temperature.

*Acknowledgements*

AP acknowledges support from the Indian Institute of Science under an IOE scheme. Further, AP and AN acknowledge DST SERB IRHPA [IPA/2020/000033] for additional support. AN additionally acknowledges support from the Indian Institute of Science start-up grant (SRG/MHRD-19-0001). PM thanks MHRD, GoI for a Prime Minister's Research Fellowship. We thank Prof. Vivek Tiwari and Prof. Satish Patil for helpful discussions regarding photophysics of molecular species.